\documentclass[conference]{IEEEtran}
\IEEEoverridecommandlockouts
\usepackage{cite}
\usepackage{amsmath,amssymb,amsfonts}
\usepackage{algorithmic}
\usepackage{graphicx}
\usepackage[numbers]{natbib}
\usepackage{textcomp}
\usepackage{xcolor}
\usepackage{listings}
\usepackage{courier}
\usepackage{url}
\usepackage[hidelinks]{hyperref}
\usepackage{relsize,xspace}
\usepackage{fancyhdr}

\def\BibTeX{{\rm B\kern-.05em{\sc i\kern-.025em b}\kern-.08em
    T\kern-.1667em\lower.7ex\hbox{E}\kern-.125emX}}
\begin{document}

\title{Blockchain-Oriented Software Variant Forks:\\ A Preliminary Study}

\author{\IEEEauthorblockN{Henrique Rocha}
\IEEEauthorblockA{\textit{Department of Computer Science} \\
\textit{Loyola University Maryland}\\
Baltimore, USA \\
henrique.rocha@gmail.com}
\and
\IEEEauthorblockN{John Businge}
\IEEEauthorblockA{\textit{Department of Computer Science / ANSYMO} \\
\textit{University of Antwerp}\\
Antwerp, Belgium \\
john.businge@uantwerpen.be}
}


\maketitle

\begin{abstract}
In collaborative social development platforms such as GitHub, forking a repository is a common activity. A variant fork wants to split the development from the original repository and grow towards a different direction. In this preliminary exploratory research, we analyze the possible reasons for creating a variant fork in blockchain-oriented software. By collecting repositories in GitHub, we created a dataset with repositories and their variants, from which we manually analyzed 86 variants. Based on the variants we studied, the main reason to create a variant in blockchain-oriented software is to support a different blockchain platform (65\%). 
\end{abstract}


\section{Introduction}

Due to the popularity of cryptocurrencies~\cite{dziembowski15}, such as BitCoin~\cite{bitcoin09} and Ether~\cite{ether14}, blockchain platforms have also become more popular~\cite{porru17, luu16}. Initially designed as a distributed ledger, nowadays blockchain has grown to address many different scenarios. For example, blockchain can be used in supply-chain~\cite{zhao21, mann18}, voting~\cite{mccorry21, killer20}, vehicular ad-hoc networks~\cite{leiding16}, and others~\cite{zeng19, bornelus20}.

Since there are many applications for blockchain~\cite{zhao21, mann18, mccorry21, killer20, leiding16, zeng19, bornelus20, liu21}, with the rising popularity of blockchain platforms there is also an increase in the development of blockchain-oriented software (BOS)~\cite{porru17,pinna21}. 

Many of these BOS are developed using coding platforms such as GitHub.
The advent of social coding platforms like GitHub, BitBucket, and GitLab have substantially improved software reuse through forking.
Developers may fork a \textit{mainline repository} into a new \textit{forked repository} and take governance over the latter while preserving the full revision history of the former. 
The social coding platforms comprise a number of \textit{software ecosystems} i.e., large collections of interdependent software components that are maintained by large and geographically distributed communities of collaborating contributors~\cite{decan:2017:saner}. 
A \textit{software family} is subset of the a \textit{software ecosystem}. A software family comprises two or more software projects sharing a common code base as well as variable code. An example of a family is a variant fork and the original project (e.g., \texttt{MariaDB} is a  variant fork of the upstream project \texttt{MySQL}). As opposed to the very common bug-fixing/feature \textit{social forks} that are usually integrated back in the upstream, \textit{variant forks} are very rare and usually created to diverge away from the upstream with no intention to contribute back~\cite{businge:emse:2021,zhou20}. 

Many cryptocurrency applications inherit code from the mainline project.
For example, the variant fork repository \href{https://github.com/PIVX-Project/PIVX}{\texttt{PIVX-Project/PIVX}} is a cryptocurrency project hosted on Github that was forked from \href{https://github.com/dashpay/dash}{\texttt{dashpay/dash}}. Moreover, \href{https://github.com/dashpay/dash}{\texttt{dashpay/dash}} was also forked from the mainline \href{https://github.com/bitcoin/bitcoin}{\texttt{bitcoin/bitcoin}}. The downstream applications continuously monitor their immediate upstream and others in the hierarchy for important updates like bug and security fixes as well as other specific updates~\cite{businge:saner:2022}. 
These three cryptocurrency BOS can be considered as a software family comprising three variants (a mainline and two variant forks)  
and all other related cryptocurrency applications form software ecosystem~\cite{businge:emse:2021, Decan:2019:emse}.

In this short paper, we conduct a preliminary study on BOS repositories with variant forks. We analyze the characteristics of the repositories and find the reason why the variant fork was created. For example, someone may fork a Wallet application to support a different blockchain platform than the original. For this study, we collected 124 mainline BOS repositories in GitHub, and 237 variant forks, for a total of 361 repositories. From those, we manually analyzed 86 variant forks which led to discovering the most common reason for creating a variant in BOS is to support another blockchain platform (65\% of the investigated cases).

The remainder of this paper is organized as follows. Section~\ref{sec:dataset} describes our collection process for the GitHub repositories which we use in this study. Section~\ref{sec:results} presents our preliminary results.
Section~\ref{sec:related-work} discuss some related work research. 
Finally, in Section~\ref{sec:conclusion}, we present our final remarks and outline future research goals and challenges. 

\section{Data Collection}\label{sec:dataset}

To collect the repositories, we employed a similar method adopted by another research that studied variants and software families~\cite{businge:emse:2021, businge:saner:2022}.
We started by looking for mainline projects (i.e., original repositories) using the Github's Rest API v3 search endpoint~\footnote{https://docs.github.com/en/rest/reference/search}. 
We identify the repository as being a BOS if in their GitHub name\,/\,description\,/\,readme contains the following keywords: \texttt{blockchain}, \texttt{ethereum}, \texttt{bitcoin}, or \texttt{cryptocurrency}. 
We then filtered out \emph{popular} ($>5$ stars and forks), \emph{long-lived} (created before 2018) and \emph{active} (still updated in 2020) repositories.
Then, for all the mainline projects we found, we tried to identify and collect variant forks. 
This process is subject to a known threat to validity since previous studies revealed that the majority of forks on GitHub are inactive~\cite{businge:2019Saner,Businge:2017} or are social forks~\cite{businge:2018icsme}.
To reduce this threat, 
we filtered forks based on the following heuristics: $\geq 3$ stars, $\geq 3$ commits ahead of the mainline. A fork with stars implies that it is liked by other persons other than developers. Commits ahead of the mainline implies that the fork has added extra functionality that is not in the original project. The is no criteria for the thresholds of the stars and commits, however, the lower numbers results in a lot of noise and higher numbers results missing possibly interesting cases. To reduce on the noise, we further filtered the mainline--variant fork pairs by ensuring they had
diverging \textsf{README} files presenting BOS applications (manually verified).
Based on the criteria detailed above, we gathered 124 mainline repositories, and 237 variant forks. A spreadsheet with the dataset is publicly available.\footnote{\url{https://bit.ly/3GV2PeO}}

\section{Preliminary Results}\label{sec:results} 

We start our analysis by looking at the number of variant forks that each mainline repository has. From the 124 mainline repositories, in 91 of them (72\%) we only identified one variant. Therefore, the median value for number of variants is one, in our collected data. The average number of variants is 1.91. Table~\ref{tab:variant_count} shows the top-ten mainline BOS repositories with the most number of variants as identified by us. 

\begin{table}[ht]
\caption{Top-10 BOS Mainlines according to the number of Variants.}
\centering
\begin{tabular}{l p{2cm} c}
\hline
\textbf{Repository}         & \textbf{Domain} & \textbf{\# of Variants} \\ \hline
\href{https://github.com/ethereum/go-ethereum}{ethereum/go-ethereum}    & Protocol / Node & 19     \\ 
\href{https://github.com/Uniswap/uniswap-interface}{Uniswap/uniswap-interface}   & Protocol / Token Exchange & 15 \\ 
\href{https://github.com/bitcoin/bitcoin}{bitcoin/bitcoin}             & Protocol / Node / Crytpcurrency & 12  \\ 
\href{https://github.com/monero-project/monero}{monero-project/monero}       & Cryptocurrency & 9 \\ 
\href{https://github.com/ethereum/solidity}{ethereum/solidity}           & Compiler & 6 \\ 
\href{https://github.com/MetaMask/metamask-extension}{MetaMask/metamask-extension} & Browser Add-on & 6 \\ 
\href{https://github.com/EOSIO/eosio.contracts}{EOSIO/eosio.contracts}       & Smart Contracts & 5 \\ 
\href{https://github.com/tendermint/tendermint}{tendermint/tendermint}       & BFT Consensus & 5 \\ 
\href{https://github.com/turtlecoin/turtlecoin}{turtlecoin/turtlecoin} & Cryptocurrency &  5 \\ 
\href{https://github.com/bitcoinj/bitcoinj}{bitcoinj/bitcoinj}& Library/Protocol & 4 \\ 
\hline 
\end{tabular}
\label{tab:variant_count}
\end{table}

We like to note that the mainline repositories we gathered (exemplified in Table~\ref{tab:variant_count}) have more forks than the ones used in this study. For instance, \href{https://github.com/ethereum/go-ethereum}{\texttt{ethereum/go-ethereum}} repository, has more than 12 thousand forks. The 19 forks we showed for this repository are the ones we were able to identify as variants.  

We manually analyze the mainline--variant fork pairs, by comparing the repository descriptions and README files, to find the possible reason for variant creation. We discarded pairs where the reason variant creation is not clear. We will analyze the forks for the top-ten mainline repositories we presented in Table~\ref{tab:variant_count}.

Table~\ref{tab:variants1} shows the variant forks we identified for the  \href{https://github.com/ethereum/go-ethereum}{\texttt{ethereum/go-ethereum}} repository. The ahead by column shows how many commits the variant is ahead of the original repository. This table also shows the possible reason for the variant creation, based on our manual analysis.

\begin{table}[ht]
\caption{Variants for ethereum/go-ethereum.}
\centering
\begin{tabular}{p{3.0cm} c p{2.2cm} }
\hline
\textbf{Variant} & \textbf{Ahead by} & \textbf{Reason} \\
\hline
\href{https://github.com/tomochain/tomochain-v1}{tomochain/tomochain-v1} & 1595 & Support a different blockchain. \\ 
\href{https://github.com/ShyftNetwork/go-empyrean}{ShyftNetwork/go-empyrean} & 748 & Support a different blockchain. \\
\href{https://github.com/expanse-org/go-expanse}{expanse-org/go-expanse} & 532 & Support a different blockchain. \\
\href{https://github.com/XinFinOrg/XDPoSChain}{XinFinOrg/XDPoSChain} & 508 & Support a different blockchain. \\
\href{https://github.com/ubiq/go-ubiq}{ubiq/go-ubiq} & 320 & Support a different blockchain. \\
\href{https://github.com/NoteGio/go-ethereum}{NoteGio/go-ethereum}& 268 & Unspecified. \\
\href{https://github.com/EthereumCommonwealth/go-callisto}{EthereumCommonwealth/go-callisto} & 150 & Support a different blockchain.\\
\href{https://github.com/second-state/lityvm}{second-state/lityvm} & 122 & Extending the protocol. \\
\href{https://github.com/Ethersocial/go-ethersocial}{Ethersocial/go-ethersocial} & 103 & Support a different blockchain. \\
\href{https://github.com/BrightID/IDChain}{BrightID/IDChain} & 47 & Support a different blockchain. \\
\href{https://github.com/blocknative/go-ethereum}{blocknative/go-ethereum} & 36 & Unspecified. \\
\href{https://github.com/nebulaai/nbai-node}{nebulaai/nbai-node} & 28 & Support a different blockchain. \\
\href{https://github.com/AlexeyAkhunov/turbo-geth-archive}{AlexeyAkhunov/turbo-geth-archive} & 26 & Efficiency improvement. \\
\href{https://github.com/Onther-Tech/go-ethereum}{Onther-Tech/go-ethereum} & 22 & Unspecified. \\
\href{https://github.com/ewasm/go-ethereum}{ewasm/go-ethereum} & 19 & Support another technology. \\
\href{https://github.com/flashbots/mev-geth}{flashbots/mev-geth} & 11 & Support another technology. \\
\href{https://github.com/Giulio2002/go-ethereum}{Giulio2002/go-ethereum} & 11 & Unspecified. \\
\href{https://github.com/loomnetwork/go-ethereum}{loomnetwork/go-ethereum} & 9 & Unspecified. \\
\href{https://github.com/eth4nos/go-ethereum}{eth4nos/go-ethereum} & 4 &  Efficiency improvement. \\
\hline 
\end{tabular}
\label{tab:variants1}
\end{table}

The \href{https://github.com/ethereum/go-ethereum}{\texttt{ethereum/go-ethereum}} is the official Go implementation of the Ethereum~\cite{ether14} protocol and its client node.
The most popular reason for the variant forks is to support a different blockchain platform (eight out of 19 forks). Since the original repository contains the core of the Ethereum implementation, it is possible that some of those variants to be sidechains\footnote{A sidechain is a separate blockchain that runs in parallel to another blockchain platform (referred as mainnet). The sidechain is connected to the mainnet and provides a bridge for transactions between both blockchains. $<$\url{https://ethereum.org/en/developers/docs/scaling/sidechains/}$>$} as well.
We also found two forks that specified some sort of efficiency or performance improvement (e.g., one is a lightweight client node for Ethereum). Moreover, we found two forks to support another technology but unrelated to the blockchain. For instance, \href{https://github.com/ewasm/go-ethereum}{\texttt{ewasm/go-ethereum}} adds support to Ewasm, which is the primary candidate to replace EVM (Ethereum Virtual Machine) as part of the Ethereum 2.0 "Serenity" roadmap.
Just one variant we found was to extend the Ethereum protocol. 
From the 19 forks, three did not specify in their description or readme file any difference from the original repository. Based on their commits, we flagged them as variants, but we are unable to find the reason for their creation.

The repository \href{https://github.com/Uniswap/uniswap-interface}{\texttt{Uniswap/uniswap-interface}} is an open-source interface for Uniswap which is a protocol for decentralized exchange of Ethereum tokens. It shares a similar domain of \href{https://github.com/ethereum/go-ethereum}{\texttt{ethereum/go-ethereum}}, but it serves a different purpose. Tokens (or Non-fungible Tokens) are very popular in Ethereum and, by consequence, so are token exchange contracts and websites. From the 15 variants, 13 are just to support a different exchange protocol or contract. The fork \href{https://github.com/NavidGoalpure/persian-uniswap-interface} {\texttt{NavidGoalpure/persian-uniswap-interface}} had the same basic functionality as the original but it was internationalized and and support for another language was added. The other remaining fork was not accessible anymore.       

The third repository, in Table~\ref{tab:variant_count}, is \href{https://github.com/bitcoin/bitcoin}{\texttt{bitcoin/bitcoin}} which is an open source BitCoin~\cite{bitcoin09} implementation of its protocol, client node, and cryptocurrency. Therefore it is very similar to \href{https://github.com/ethereum/go-ethereum}{\texttt{ethereum/go-ethereum}}. 
Ten out of the 12 forks are to support a different blockchain platform and create a new cryptocurrency. This is a much higher amount of forks to support another platform than the Ethereum variants. Probably, because the Ethereum repository has other functionalities, while the BitCoin one is more limited and tied to its cryptocurrency. The remaining two forks are extensions to the regular BitCoin. For instance, \href{https://github.com/OmniLayer/omnicore}{\texttt{OmniLayer/omnicore}} adds a communication layer to enable more sophisticated smart contracts in Bitcoin; and \href{https://github.com/jlopp/statoshi}{\texttt{jlopp/statoshi}} claims to bring more transparency to the nodes.

In Table~\ref{tab:variants4}, we show the variant forks for Monero (\href{https://github.com/monero-project/monero}{\texttt{monero-project/monero}}), which is a cryptocurrency. We were expecting similar results from \href{https://github.com/bitcoin/bitcoin}{\texttt{bitcoin/bitcoin}}, as most variant forks for a cryptocurrency are probably to create another type of cryptocurrency, which is similar in principle to support a new blockchain platform. 

\begin{table}[ht]
\caption{Variants for {monero-project/monero}.}
\centering
\begin{tabular}{l c p{2.2cm} }
\hline
\textbf{Variant} & \textbf{Ahead By} & \textbf{Reason} \\
\hline
\href{https://github.com/oxen-io/oxen-core}{oxen-io/oxen-core} & 3894 & Support a different Cryptocurrency\\
\href{https://github.com/Beldex-Coin/beldex}{Beldex-Coin/beldex} & 1034 & Support a different Cryptocurrency\\
\href{https://github.com/aeonix/aeon}{aeonix/aeon} & 980 & Support a different Cryptocurrency\\
\href{https://github.com/electroneum/electroneum}{electroneum/electroneum} & 808 & Support a different Cryptocurrency\\
\href{https://github.com/EquilibriaCC/Equilibria}{EquilibriaCC/Equilibria} & 558 & Support a different Cryptocurrency\\
\href{https://github.com/X-CASH-official/xcash-core}{X-CASH-official/xcash-core} & 420 & Support a different Cryptocurrency\\
\href{https://github.com/swap-dev/swap}{swap-dev/swap} & 129 & Support a different Cryptocurrency\\
\href{https://github.com/dweab/haven-do-not-use}{dweab/haven-do-not-use} & 103 & Support a different Cryptocurrency\\
\href{https://github.com/toints/moneroclassic}{toints/moneroclassic} & 45 & Maintain the classic version of the original cryptocurrency.\\
\hline 
\end{tabular}
\label{tab:variants4}
\end{table}

As we expected, the majority of the variant forks for \href{https://github.com/monero-project/monero}{\texttt{monero-project/monero}} are to support or create another type of cryptocurrency (eight out of nine forks). There is one fork that is used to maintain and support the classic version of Monero. 

The fifth repository (according to Table~\ref{tab:variant_count}) is \href{https://github.com/ethereum/solidity}{\texttt{ethereum/solidity}} which contains the language definition, documentation, and compiler for Solidity~\cite{solidity21}. Solidity is one of the major languages for coding smart contracts in Ethereum. From its variants (Table~\ref{tab:variants5}), two forks are extensions for the compiler (model checking, and static verifier). One variant is to support a different platform, by compiling Solidity into Tron Virtual Machine. Another fork is a translation for the documentation of Solidity. Finally, there were two forks we could not identify the reason for their creation. 

\begin{table}[ht]
\caption{Variants for {ethereum/solidity}.}
\centering
\begin{tabular}{l c p{2.2cm} }
\hline
\textbf{Variant} & \textbf{Ahead By} & \textbf{Reason} \\
\hline
\href{https://github.com/SRI-CSL/solidity}{SRI-CSL/solidity} & 1348 & Compiler Extension  \\ 
\href{https://github.com/ScottWe/solidity-to-cmodel}{ScottWe/solidity-to-cmodel} & 364 & Compiler extension  \\ 
\href{https://github.com/tronprotocol/solidity}{tronprotocol/solidity} & 174 & Support a different platform \\ 
\href{https://github.com/akira-19/solidity}{akira-19/solidity} & 68 & Unspecified \\ 
\href{https://github.com/Karocyt/solidity-fr}{Karocyt/solidity-fr} & 62 & Translation \\ 
\href{https://github.com/PlatONnetwork/solidity}{PlatONnetwork/solidity} & 4 & Unspecified \\ 
\hline 
\end{tabular}
\label{tab:variants5}
\end{table}

In Table~\ref{tab:variant_count}, the sixth repository with most variants is \href{https://github.com/MetaMask/metamask-extension}{\texttt{MetaMask/metamask-extension}} where we identified six forks (Table~\ref{tab:variants6}). MetaMask is a browser extension (Chrome and Firefox) that facilitates browsing Ethereum blockchain on certain websites. This repository is from a different domain than the previous ones, by being a browser add-on.

\begin{table}[ht]
\caption{Variants for {MetaMask/metamask-extension}.}
\centering
\begin{tabular}{l c p{2.2cm} }
\hline
\textbf{Variant} & \textbf{Ahead By} & \textbf{Reason} \\
\hline
\href{https://github.com/poanetwork/nifty-wallet}{poanetwork/nifty-wallet} & 1480 & Enhancement \\ 
\href{https://github.com/Conflux-Chain/conflux-portal}{Conflux-Chain/conflux-portal} & 869 & Support a different blockchain \\
\href{https://github.com/smilofoundation/SmiloWallet-extension}{smilofoundation/SmiloWallet-extension} & 142 & Support a different blockchain \\
\href{https://github.com/ubiq/sparrow-extension-OLD}{ubiq/sparrow-extension-OLD} & 90 & Support a different blockchain \\
\href{https://github.com/CoboVault/metamask-extension}{CoboVault/metamask-extension} & 64 & Unspecified \\
\href{https://github.com/dsrvlabs/celo-extension-wallet}{dsrvlabs/celo-extension-wallet} & 29 & Support a different blockchain \\
\hline 
\end{tabular}
\label{tab:variants6}
\end{table}

Four out of six MetaMask variants were to support a different blockchain platform. One fork was a usability enhancement for the add-on. And one fork, we could not discover the reason for its creation.

The seventh repository (Table~\ref{tab:variant_count}) we looked was \href{https://github.com/EOSIO/eosio.contracts}{\texttt{EOSIO/eosio.contracts}} with five variants. This repository contains smart contracts to provide basic functions of the EOSIO blockchain.
Four out of five variants were created to support another blockchain platform that uses EOSIO as a basis. One fork did not specify its purpose on the description or readme file.

The eighth repository (Table~\ref{tab:variant_count}) is a middleware for the application-based blockchain platform that uses a Byzantine Fault Tolerant (BFT) consensus algorithm, Tendermint~\cite{amoordon19} (\href{https://github.com/tendermint/tendermint}{\texttt{tendermint/tendermint}}). Since application-based blockchains (such as Tendermint, and Hyperledger Fabric) have major differences from public blockchain platforms (such as Ethereum, and BitCoin), we were expecting to observe different reasons for the creation of variants. However, four of the five variants did not specify anything to justify their creation, and the description and readme files were identical to the original repository. One fork, \href{https://github.com/QuarkChain/tendermintx}{\texttt{QuarkChain/tendermintx}}, was created as an extension to provide greater flexibility for Tendermint. 

The crytocurrency Turtlecoin is the ninth repository (\href{https://github.com/turtlecoin/turtlecoin}{\texttt{turtlecoin/turtlecoin}}) in number of variant forks.  
All five variants we found for this repository are to support or create a new cryptocurrency based on the Turtlecoin original code. 

Finally, the final repository on our top-ten list with most variants (Table~\ref{tab:variant_count}) is the \href{https://github.com/bitcoinj/bitcoinj}{\texttt{bitcoinj/bitcoinj}}. BitCoinJ is a Java library that implements the BitCoin protocol, has a built-in wallet, and can communicate with the BitCoin blockchain without the need for any other external code. Three out of the four forks were created to support a different blockchain platform and cryptocurrency than BitCoin. One fork does not specify the reason for its creation. \\

\textbf{Summarizing.}
We aimed to identify the possible reason for the creation of BOS variants by analyzing the top-ten mainline repositories with most variant forks. In total, we manually investigated 86 variant forks from those top-ten repositories. We can see the most popular reason for creating a variant in BOS is to support a different blockchain platform (65\% of the variants or 56 out of 86). The second most popular reason is the extend the original with additional features or technologies (10\% or nine out of 86). 
Approximately 14\% of the variants (12 out of 86) did not specify any difference with the main repository in their description or readme file. 

\section{Related Work}\label{sec:related-work}

Most studies analysed forking on Sourceforge, pre-dating the advent of social coding platforms like GitHub~\cite{Linus:2012Perspectives,Gregorio:2012}. Several of those early studies report perceived controversy around variant forks~\cite{Chua:Forking:2017}. Jiang et al.~\cite{Lo:2017} state that, although forking may have been controversial in the open source software (OSS) community, it is now encouraged and a built-in feature on GitHub.

We have encountered few studies analysing variant forks in the social coding era~\cite{businge:2018icsme,zhou20,businge20,businge:emse:2021}.
Businge et al.~\cite{businge:2018icsme} study focused on the Android ecosystem and found that re-branding, simple customizations, feature extension, and implementation of different but related features are the main motivations to create a fork for Android apps.
Zhou et al.~\cite{zhou20} interviewed 18 developers of hard forks on GitHub to understand reasons for forking in modern social coding environments that explicitly support forking.
Businge et al.~\cite{businge:emse:2021} investigated the interaction between mainline and variants. The authors quantitatively investigated code propagation among variants and their mainline in three software ecosystems. They found that only about 11\% of the 10,979 mainline--variant pairs had integrated code among themselves.

To the best of our knowledge, there are no studies investigating variant forks for blockchain-oriented software. 

\section{Final Remarks}\label{sec:conclusion}

In this short paper, we presented a preliminary exploratory research on the possible reasons for creating a variant fork in blockchain-oriented software. We collected repositories from GitHub and manually analyzed 86 variant forks from the top-ten mainlines in our dataset. 

We discovered in our dataset that for BOS variants, 65\% were created to support a different blockchain platform. This seems like a reasonable and fast way to develop software for a specific blockchain platform. Instead of coding everything from scratch, a developer finds a similar BOS repository that accomplishes its main goal, forks it, and then just changes the protocol to interact with a different blockchain. However, before our study, we did not know how common this practice is.

The second most common reason for the creation of a variant is to extend the mainline with additional features or technologies, occurring in 10\% of the analyzed forks. Why these extensions were not integrated into the original repository remains an open question.

For future research, we plan to investigate all the variants in our current dataset. Additionally, we will analyze other aspects of the variants compared to the mainline such as number of commits, commit size, community size, number of stars, technical debt, etc. We also plan to contrast our findings in BOS repositories with non-blockchain software to properly compare the differences between them. Moreover, we would like to contact the main developers or owners from the variants to know the reasons why the variant was created. However, since GitHub policy rules prohibit sending unsolicited emails, we are not sure how to contact those developers.

\section*{Acknowledgment}
John Businge's work is supported by the FWO-Vlaanderen and F.R.S.-FNRS via the EOS project 30446992 SECO-ASSIST.

\bibliographystyle{IEEEtranN}
\bibliography{references}

\end{document}